%% file: paper.tex
\documentclass[twocolumn,showpacs,aps,prl,superscriptaddress]{revtex4} 

\usepackage{graphicx}
\usepackage{dcolumn}
\usepackage{amsmath}
\usepackage{xspace}
\usepackage{float}
\usepackage{epsfig}

\input babarsym
\newcommand{\BABARPubYear}    {07}
\newcommand{\BABARPubNumber}  {033}
\newcommand{\SLACPubNumber}   {12594}

\newcommand{\PRLeeg}{\ensuremath{\epem\gamma}\xspace}
\newcommand{\PRLmmg}{\ensuremath{\mumu\gamma}\xspace}
\newcommand{\PRLresultE}{1.2\xspace}
\newcommand{\PRLresultM}{1.5\xspace}
\newcommand{\PRLnObsE}{1\xspace}
\newcommand{\PRLnObsM}{1\xspace}
\newcommand{\PRLnBgE}{\ensuremath{1.75\pm 1.38\pm 0.36}\xspace}
\newcommand{\PRLnBgM}{\ensuremath{2.66\pm 1.40\pm 1.58}\xspace}
\newcommand{\PRLsigeffE}{\ensuremath{7.4\pm 0.3}\xspace}
\newcommand{\PRLsigeffM}{\ensuremath{5.2\pm 0.2}\xspace}
\newcommand{\PRLnULE}{2.82\xspace}
\newcommand{\PRLnULM}{2.55\xspace}

\long\def\inst#1{\par\nobreak\kern 4pt\nobreak
    {\it #1}\par\vskip 10pt plus 3pt minus 3pt}

\begin{document}

\begin{flushleft}
  \babar-PUB-\BABARPubYear/\BABARPubNumber\\
  SLAC-PUB-\SLACPubNumber\\
\end{flushleft}

\title{
 Search for the Decays
 \boldmath{\beeg} and \boldmath{\bmmg}
}
\input pubboard/authors_may2007.tex
\date{\today}

\begin{abstract}
We present results of a search for the decays \bllg ($\ell=e$, $\mu$).  The search is performed using \PRLnBB \BB pairs collected at the \FourS resonance with the \babar\ detector at the \pep2 \B Factory at SLAC. We find no significant signal and set the following branching fraction upper limits at the 90\% confidence level: $\BR(\beeg)<\PRLresultE\times 10^{-7}$ and $\BR(\bmmg)<~\PRLresultM\times 10^{-7}$.
\end{abstract}
 
\maketitle
\par    

\begin{figure*}[!htb]
\begin{center}
\includegraphics[height=3cm]{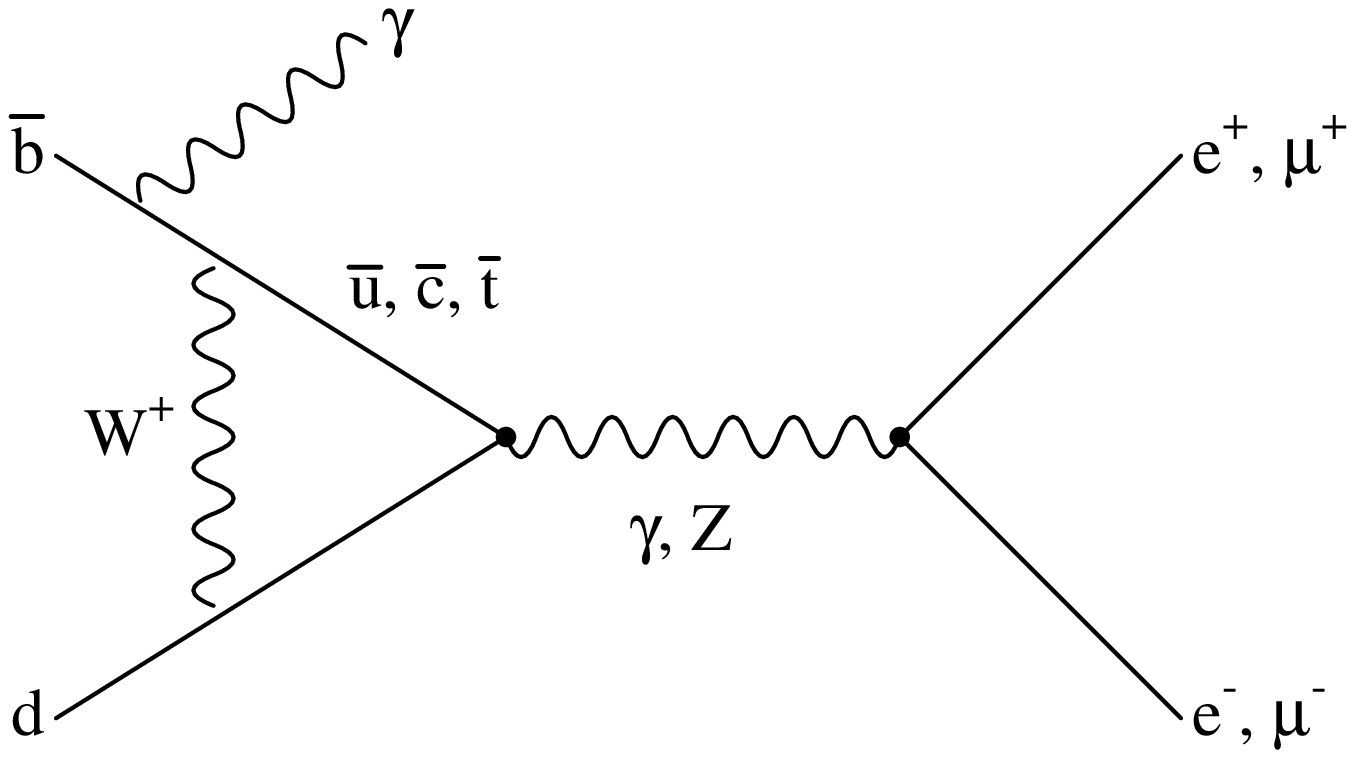}\hspace{0.1 in}
\includegraphics[height=3cm]{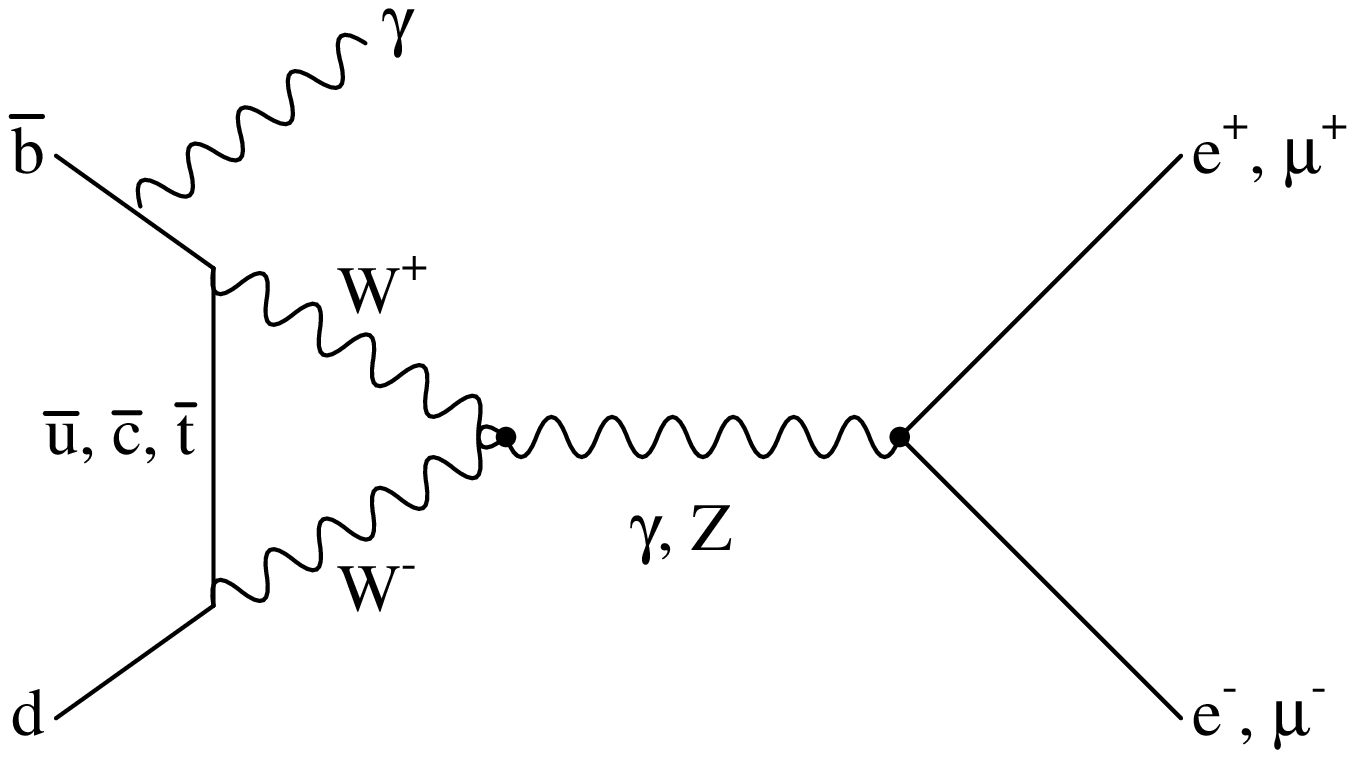}\hspace{0.1 in}
\includegraphics[height=3cm]{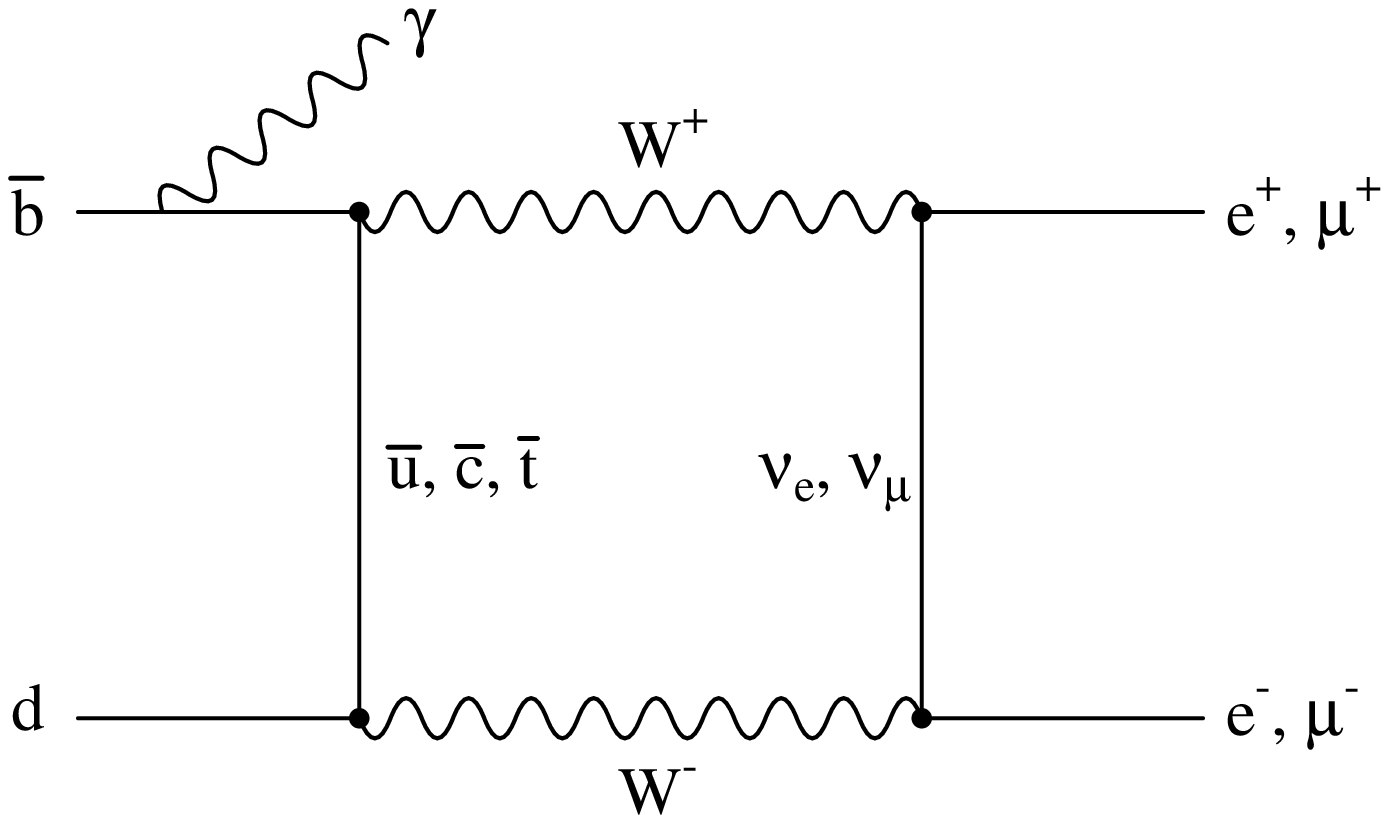}
\caption{The penguin (left and middle) and box (right) Feynman diagrams for \bllg ($\ell=e,\mu$) decays. The photon can be emitted from any of the quarks or leptons, but the amplitudes are largest if the photon is emitted from one of the initial quarks.}
\label{fig:feyn}
\end{center}
\end{figure*}

Radiative leptonic decays of neutral \B mesons, \bllg with $\ell=e,$ $\mu$~\cite{bib:cc}, are flavor-changing neutral-current transitions that are forbidden at the tree level in the standard model (SM). In the SM, such processes are described by penguin and box diagrams to leading order, as shown in \PRLfig~\ref{fig:feyn}. The expected branching fractions for these processes are of order of $10^{-10}$~\cite{bib:SM}. A measured branching fraction bigger than this would be an evidence for new physics. The largest contributions arise from processes in which a photon is emitted from one of the initial quarks, thus avoiding the helicity suppression of the purely leptonic decay \Bz\to\ellell.
A search for the processes \Bz\to\ellell has been performed by the \babar\ collaboration and others~\cite{bib:B2LL}, but there is no previous search for the \bllg decays.

The analysis described in this Letter uses a sample of \PRLnBB \BB pairs recorded with the \babar\ detector at the \pep2 asymmetric energy \epem storage rings. This corresponds to an integrated luminosity of 292\invfb collected at the \FourS resonance. 

A detailed description of the \babar\ detector can be found elsewhere~\cite{bib:babar}. Charged-particle trajectories are measured by a five-layer silicon vertex tracker and a 40-layer drift chamber operating in a 1.5 T magnetic field. A detector of internally reflected Cherenkov light is used for charged hadron identification. Surrounding this is an electromagnetic calorimeter (EMC) consisting of 6580 CsI(Tl) crystals, and the instrumented flux return for the solenoid, which consists of layers of steel interspersed with resistive plate chambers or limited streamer tubes.

 A full \babar\ \MC (MC) simulation using \geantFour~\cite{bib:geant4} is used to evaluate signal efficiencies and to identify and study background sources. The signal MC sample is based on a calculation where the \bllg transition depends on three Wilson coefficients $C_7$, $C_9$, and $C_{10}$ at leading order~\cite{bib:signalMC}.

We reconstruct the \Bz signal candidates by combining two oppositely-charged leptons and a photon. The \Bz vertex is fitted using a Kalman Filter method~\cite{bib:kalman}. The leptons are required to originate from a common vertex, and the \Bz candidate is required to be consistent with coming from the beam interaction point.

To minimize the number of misidentified particles, the leptons are required to satisfy stringent particle identification criteria~\cite{bib:PID}. For the electron candidates, the energy loss due to bremsstrahlung is recovered whenever possible, by looking for the energy deposits (clusters) in the EMC close to the intersection of their tracks with the EMC. For photon clusters, the transverse shower shape is required to be consistent with an electromagnetic shower. Leptons and photons are required to reside fully in the geometric acceptance of the detector.

Since the signal event contains two neutral \B mesons and no additional particles, the total energy of each \B meson in the center-of-mass (CM) frame must be equal to half of the total beam energy in the CM frame. We define $\mes=\sqrt{(E_{\text{beam}}^*)^2-(\sum_i{\bf p}_i^*)^2}$ and $\dE=\sum_i\sqrt{m_i^2+({\bf p}_i^*)^2}-E_{\text{beam}}^*$, where $E_{\text{beam}}^*$ is the beam energy in the \hjCM frame, ${\bf p}_i^*$ and $m_i$ are the momenta in the \hjCM frame and the masses of the daughter particles $i$ ($i=\ell^+,\ell^-,\gamma$), respectively. $E_{\text{beam}}^*$ is used instead of the measured \B meson energy in the \hjCM frame because $E_{\text{beam}}^*$ is more precisely known. For correctly reconstructed \Bz mesons, the \mes distribution has a maximum at the \Bz mass with a standard deviation of about 3\mevcc and the \dE distribution has a maximum near zero with a standard deviation of about 30\mev.

\begin{figure}[!htb]
\begin{center}
\vspace{0.05 in}
\includegraphics[height=5cm]{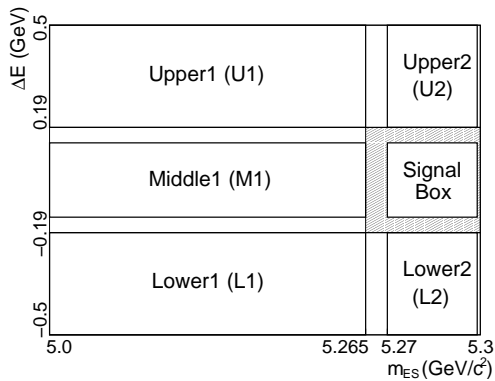}\hspace{0.1 cm}
\caption{Definitions of the signal box, blinded area (equal to the sum of the signal box and of the shaded area), and sideband areas in the \dE vs. \mes plot: Upper1 (U1), Upper2 (U2), Lower1 (L1), Lower2 (L2), and Middle1 (M1). The signal box has the same \dE range as the M1 box (different for each mode), and the same \mes limits as the U2 and L2 boxes. The figure is not drawn to scale.}
\label{fig:box}
\end{center}
\end{figure}

The \bllg candidates are selected by requiring $-0.5\leq\dE\leq 0.5\gev$ and $5.0\leq\mes\leq 5.3\gevcc$. These ranges include both background- and signal-dominated regions. As shown in \PRLfig~\ref{fig:box}, five background-dominated regions (sideband areas) are used for the background estimation. To avoid experimenter's bias, the events in the signal-dominated region (signal box) and in the shaded area covering the signal box are not included in the analysis until the final selection criteria have been optimized and the background estimation has been finalized. The shapes of the \mes and \dE distributions of the signal MC are parameterized by the Crystal Ball function~\cite{bib:CB} to allow for the asymmetric shape of the signal peak due to energy loss in the EMC. The size of the signal box is chosen to be approximately $\pm 3\times$FWHM for \dE and \mes: $-0.146(-0.112)\leq\dE\leq\ 0.082\gev$ for the \PRLeeg(\PRLmmg) mode, and $5.270\leq\mes\leq 5.289\gevcc$ for both modes.

The dominant backgrounds are: 1) unmodeled higher-order QED and hadronic two-photon processes for the \PRLeeg mode; 2) \B decays where the photon comes from a \piz decay, or the lepton is from a \jpsi or \psitwos decay; and 3) continuum background from  $\epem\rightarrow f\bar{f}$ (where $f=\u,$ $\d,$ $\s,$ $\c$, or $\tau$) processes at the parton level.

To take into account higher-order QED and hadronic two-photon processes, we introduce additional selection criteria for the \PRLeeg candidates: we require the cosine of the polar angle of $e^-$ ($\gamma$) to be between $-0.743$ ($-0.618$) and $0.81$ ($0.8$), the energy of the photon to be $\geq 0.3\gev$, the number of charged tracks (EMC clusters) in the event to be $\geq$ 5 (10), and the ratio of the second-to-zeroth order Fox-Wolfram moments (\Rtwo)~\cite{bib:Fox-Wolfram}, which is calculated with the charged tracks and neutral clusters in the rest of the event (ROE), to be $\leq 0.7$.

To reduce the number of events where the photon is from a \piz decay, we veto photon candidates that can be combined with any other photon in the event to form a \piz candidate with a mass within three standard deviations (\hjsim 20\mevcc) of the nominal \piz mass.

We veto lepton candidates that form a suitable \jpsi or \psitwos, as described in Ref.~\cite{bib:jpsiveto}.

To suppress the continuum background, we require \Rtwo, calculated from all charged tracks and neutral clusters, to be less than 0.35, and the absolute value of the cosine of the angle between the thrust axis of the \Bz candidate and that of the ROE to be less than 0.8. These variables are used in a neural network combined with the following variables: 1) the absolute value of the cosine of the angle between the \Bz direction and the beam axis, 2) the absolute value of the cosine of the angle between the thrust axis of the \Bz candidate's decay products and the beam axis, 3) the ratio of second order to zeroth order Legendre moments of all charged tracks and neutral clusters, and 4) the invariant mass of the dileptons. The neural network rejects 20(36)\% of the background while keeping 95(89)\% of the signal, for the \PRLeeg(\PRLmmg) mode. All the selection criteria are optimized with MC samples to discriminate signal from background.

After all requirements are applied, there are on average 1.01(1.07) candidates per event for the \PRLeeg(\PRLmmg) mode. In events with multiple candidates, the one with the highest probability for the vertex fit is retained. The signal efficiency is 7.4(5.2)\% for the \PRLeeg(\PRLmmg) mode. The \PRLeeg mode has higher efficiency because electrons have higher detection efficiency than muons.

To assess possible background contributions that peak in the signal box, we examined 32 exclusive hadronic and semileptonic \B decays using MC, including events where both \B mesons decay semileptonically, and found no significant contribution.

A variety of methods to estimate the background in the signal box have been tried, including fitting and counting methods in various \mes and \dE sideband areas with different conditions. All studies yield results that are compatible within uncertainties.

The chosen method is model-independent, is based on data only, and has a small systematic uncertainty. To estimate the background level in the signal box, five different sideband areas are used, as indicated in \PRLfig~\ref{fig:box}. The ratio $R^M_{est}$ is the estimated ratio of the yield in the signal box to the yield in the M1 box. The expected background in the signal box (\nbgexp) is calculated by multiplying $R^M_{est}$ by the yield in the M1 box. We estimate $R^M_{est}$ as the mean of two ratios $R^{U}$ and $R^L$, where $R^{U(L)}=N^{U2(L2)}/N^{U1(L1)}$, and where $N^X$ is the yield in box $X$. This assumes that the changes in the ratio $R^L$, $R^M_{est}$ and $R^U$ are linear in \dE.

To test our assumption of this linearity, we use MC samples and calculate the ratio $R^{M}$ by dividing the yield in the signal box by the yield in the M1 box. The relative difference between $R^M$ and $R^M_{est}$ in MC samples is assigned as a systematic uncertainty. The estimated background is \PRLnBgE (\PRLnBgM) events for the \PRLeeg(\PRLmmg) mode, where the stated errors represent the statistical and systematic uncertainties, respectively.

The dominant source of systematic uncertainty on the signal yield is the calculation used for the signal MC~\cite{bib:signalMC}. The three theoretical input parameters, the Wilson coefficients $C_7$, $C_9$, and $C_{10}$, used in the calculation are varied by \hjpm10\%, as recommended by the authors of~\cite{bib:signalMC}. This variation changes the kinematics of the signal events and can thereby impact the detection efficiency. The largest relative change in signal efficiency by this variation is assigned as a systematic uncertainty.

We have studied $\epem\rightarrow\mumu\gamma$ decays in data to assess the systematic uncertainty in photon reconstruction.

The systematic uncertainty from the lepton identification has been determined using an independent control sample of \jpsi decays.

The uncertainty on the number of \BB events is 1.1\%~\cite{bib:sys-nBB}.

The systematic uncertainty related to an imperfect detector simulation is studied using a control sample of $\Bz\rightarrow\jpsi\KS$ events. The same continuum background suppression requirements are applied on this sample and the signal efficiency is calculated. The relative difference in the signal efficiencies between data and MC samples is assigned as a systematic uncertainty.

The systematic uncertainty related to the tracking efficiency is determined from $\epem\rightarrow\tautau$ interactions, with one $\tau$ decaying leptonically and the other to three charged hadrons. All the contributions to the systematic uncertainties are added in quadrature and summarized in \PRLtab~\ref{tab:syserr}.

\begin{table}[ht!]
\caption{Summary of the systematic uncertainties on the signal yields.}
\label{tab:syserr}
\begin{center}
 \begin{tabular}{lcc} \hline
  & \PRLeeg (\%) & \PRLmmg (\%) \\ \hline \hline
 Signal Calculation      & 2.3  & 3.8 \\ 
 Photon Reconstruction   & 1.6  & 1.6 \\
 Lepton Identification   & 0.7  & 1.3 \\
 Number of \BB Pairs      & 1.1  & 1.1 \\
 Data/MC comparison       & 1.3  & 0.4 \\ 
 Tracking Efficiency      & 0.9  & 0.9 \\ \hline 
 Total                    & 3.5  & 4.6 \\ \hline 
\end{tabular}
\end{center}
\end{table}

\begin{figure}[!htb]
\begin{center}
\includegraphics[height=3.87cm]{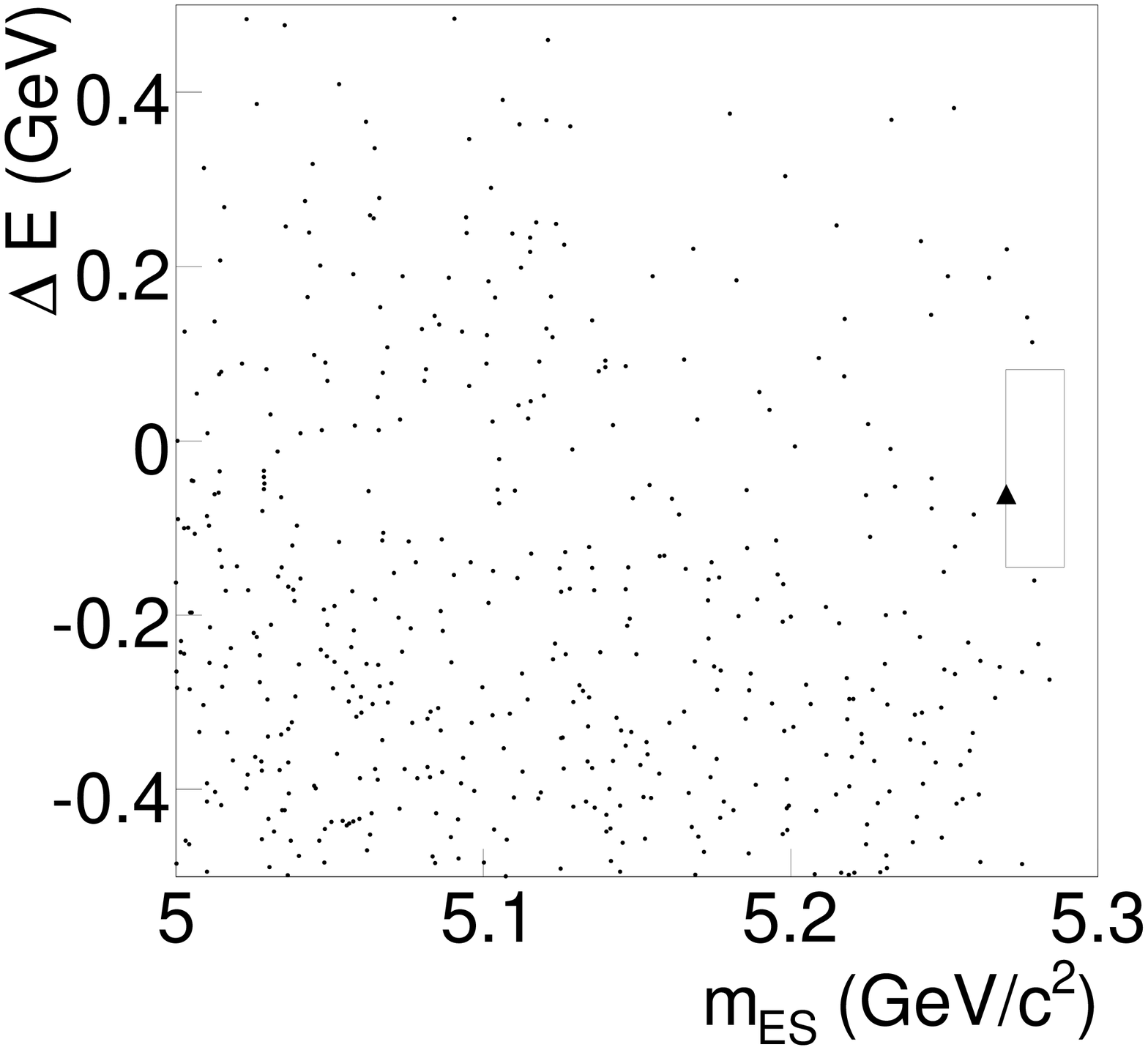}\hspace{0.1 cm}
\includegraphics[height=3.87cm]{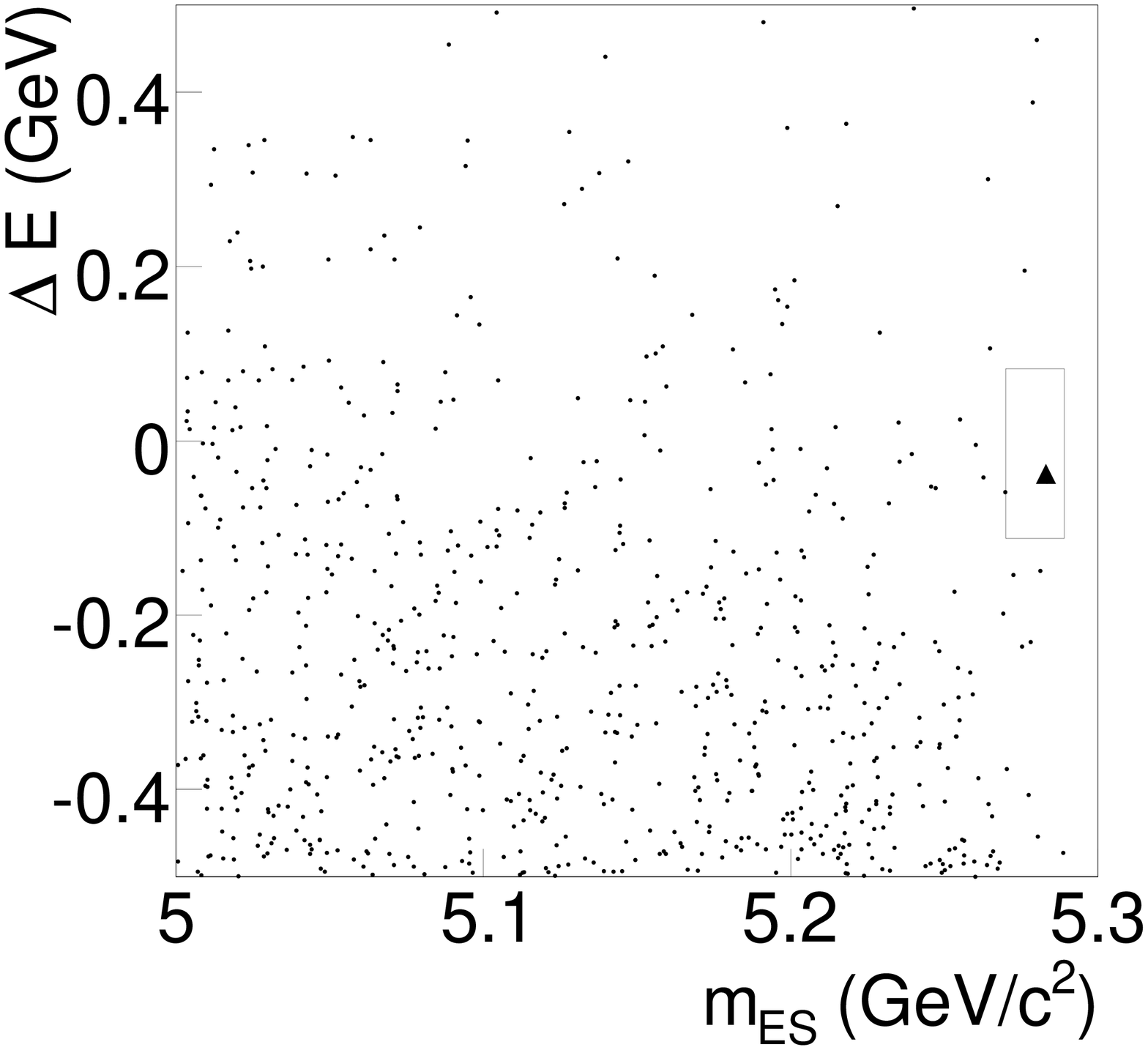}
\caption{Distribution of events in \mes and \dE. The left plot is for the \PRLeeg mode and the right plot is for the \PRLmmg mode. The dots are the events outside the signal box (rectangular region), and the triangles are the events inside the signal box. }
\label{fig:result}
\end{center}
\end{figure}

After applying the selection criteria we find one event in the signal box for each mode, as shown in \PRLfig~\ref{fig:result} and \PRLtab~\ref{tab:result}. These numbers are compatible with the expected background for both modes.

An upper limit on the branching fraction is computed from 
\begin{equation}
\BR_{UL}(\bllg)=\frac{N_{UL}}{N_{\Bz}\cdot\effsig},
\end{equation}
where $N_{UL}$ is the 90\% confidence level (C.L.) upper limit for the signal yield, determined by taking into account the one observed event in the signal box and the estimated background, using the frequentist method described in Ref.~\cite{bib:barlow} including both statistical and systematic uncertainties, $N_{\Bz}$  is the number of neutral \B mesons and \effsig is the signal reconstruction efficiency. The systematic uncertainties are included in \effsig. It is assumed that $\BR(\FourS\rightarrow\BzBzb)=\BR(\FourS\rightarrow\BpBm)$, and so $N_{\Bz}$ is equal to the number of \BB events. The 90\% C.L. branching fraction upper limits obtained are $\BR(\beeg)<\PRLresultE\times 10^{-7}$ and $\BR(\bmmg)<\PRLresultM\times 10^{-7}$.

\begin{table}[!htb]
 \caption[] {Summary of the results where \nobs (\nbgexp) is the number of observed (expected background) events in the signal box, \effsig is the efficiency, $N_{UL}$ is the 90\% C.L. upper limit for the signal yield, and $\BR_{UL}$ is the upper limit on the branching fraction at the 90\% C.L. The stated uncertainties on \nbgexp are statistical and systematic, and the uncertainty on \effsig is systematic.}
 \begin{center}
 \begin{tabular}{cccccc}  \hline
 Mode & \nobs & \nbgexp & \effsig (\%) & $N_{UL}$ & $\BR_{UL}$\\  \hline\hline
 \PRLeeg   & \PRLnObsE & \PRLnBgE & \PRLsigeffE  & \PRLnULE & $\PRLresultE \times 10^{-7}$\\
 \PRLmmg   & \PRLnObsM & \PRLnBgM & \PRLsigeffM  & \PRLnULM & $\PRLresultM \times 10^{-7}$\\ \hline
 \end{tabular}
 \end{center}
\label{tab:result}
\end{table}

In summary, a search for \bllg ($\ell=e$ or $\mu$) decays has been performed based on \PRLnBB \BB events. We obtain 90\% C.L. upper limits for the branching fractions of $\BR(\beeg)<\PRLresultE\times 10^{-7}$ and $\BR(\bmmg)<\PRLresultM\times 10^{-7}$, which represent the first limits placed on these decay channels. These are well above the SM expectations.

\input pubboard/acknow_PRL

\end{document}

%% file: babarsym.tex
\newcommand{\bllg}{\ensuremath{\Bz \rightarrow \ell^+\ell^-\gamma}\xspace}
\newcommand{\beeg}{\ensuremath{\Bz \rightarrow e^+e^-\gamma}\xspace}
\newcommand{\bmmg}{\ensuremath{\Bz \rightarrow \mu^+\mu^-\gamma}\xspace}
\newcommand{\dE}{\ensuremath{\Delta E}\xspace}
\newcommand{\mes}{\ensuremath{m_{ES}}\xspace}

\newcommand{\MC}{Monte Carlo\xspace}

\newcommand{\effsig}{\ensuremath{\epsilon_{sig}}\xspace}
\newcommand{\nbgexp}{\ensuremath{n_{bg}^{exp}}\xspace}
\newcommand{\nobs}{\ensuremath{n_{obs}}\xspace}

\newcommand{\Rtwo}{\ensuremath{R_2}\xspace}

\newcommand{\hjsim}{\ensuremath{\sim}}

\newcommand{\hjpm}{\ensuremath{\pm}}

\newcommand{\hjCM}{CM\xspace}

\newcommand{\PRLfig}{Fig.}
\newcommand{\PRLtab}{Table}
\newcommand{\hjetal}{{\it et al.}}
\newcommand{\PRLnBB}{\ensuremath{320\times 10^{6}}\xspace}

\usepackage{relsize}
\def\babar{\mbox{\slshape B\kern-0.1em{\smaller A}\kern-0.1em
    B\kern-0.1em{\smaller A\kern-0.2em R}}}

\def\epem       {\ensuremath{e^+e^-}\xspace}

\def\mumu       {\ensuremath{\mu^+\mu^-}\xspace}

\def\tautau     {\ensuremath{\tau^+\tau^-}\xspace}

\def\ellell     {\ensuremath{\ell^+ \ell^-}\xspace}

\def\u     {\ensuremath{u}\xspace}

\def\d     {\ensuremath{d}\xspace}

\def\s     {\ensuremath{s}\xspace}

\def\c     {\ensuremath{c}\xspace}

\def\piz   {\ensuremath{\pi^0}\xspace}

\def\Kbar  {\kern 0.2em\overline{\kern -0.2em K}{}\xspace}

\def\Kz    {\ensuremath{K^0}\xspace}
\def\Kzb   {\ensuremath{\Kbar^0}\xspace}
\def\KzKzb {\ensuremath{\Kz \kern -0.16em \Kzb}\xspace}
\def\Kp    {\ensuremath{K^+}\xspace}
\def\Km    {\ensuremath{K^-}\xspace}

\def\KpKm  {\ensuremath{\Kp \kern -0.16em \Km}\xspace}
\def\KS    {\ensuremath{K^0_{\scriptscriptstyle S}}\xspace}

\def\Dbar    {\kern 0.2em\overline{\kern -0.2em D}{}\xspace}

\def\Dz      {\ensuremath{D^0}\xspace}
\def\Dzb     {\ensuremath{\Dbar^0}\xspace}
\def\DzDzb   {\ensuremath{\Dz {\kern -0.16em \Dzb}}\xspace}
\def\Dp      {\ensuremath{D^+}\xspace}
\def\Dm      {\ensuremath{D^-}\xspace}

\def\DpDm    {\ensuremath{\Dp {\kern -0.16em \Dm}}\xspace}

\def\B       {\ensuremath{B}\xspace}
\def\Bbar    {\kern 0.18em\overline{\kern -0.18em B}{}\xspace}

\def\BB      {\ensuremath{B\Bbar}\xspace} 
\def\Bz      {\ensuremath{B^0}\xspace}
\def\Bzb     {\ensuremath{\Bbar^0}\xspace}
\def\BzBzb   {\ensuremath{\Bz {\kern -0.16em \Bzb}}\xspace}
\def\Bu      {\ensuremath{B^+}\xspace}
\def\Bub     {\ensuremath{B^-}\xspace}

\def\BpBm    {\ensuremath{\Bu {\kern -0.16em \Bub}}\xspace}

\def\BorBbar    {\kern 0.18em\optbar{\kern -0.18em B}{}\xspace}
\def\DorDbar    {\kern 0.18em\optbar{\kern -0.18em D}{}\xspace}
\def\KorKbar    {\kern 0.18em\optbar{\kern -0.18em K}{}\xspace}

\def\jpsi     {\ensuremath{{J\mskip -3mu/\mskip -2mu\psi\mskip 2mu}}\xspace}
\def\psitwos  {\ensuremath{\psi{(2S)}}\xspace}

\mathchardef\Upsilon="7107
\def\Y#1S{\ensuremath{\Upsilon{(#1S)}}\xspace}

\def\FourS {\Y4S}

\mathchardef\Deltares="7101
\mathchardef\Xi="7104
\mathchardef\Lambda="7103
\mathchardef\Sigma="7106
\mathchardef\Omega="710A

\def\Deltabar{\kern 0.25em\overline{\kern -0.25em \Deltares}{}\xspace}
\def\Lbar{\kern 0.2em\overline{\kern -0.2em\Lambda\kern 0.05em}\kern-0.05em{}\xspace}
\def\Sigbar{\kern 0.2em\overline{\kern -0.2em \Sigma}{}\xspace}
\def\Xibar{\kern 0.2em\overline{\kern -0.2em \Xi}{}\xspace}
\def\Obar{\kern 0.2em\overline{\kern -0.2em \Omega}{}\xspace}
\def\Nbar{\kern 0.2em\overline{\kern -0.2em N}{}\xspace}
\def\Xb{\kern 0.2em\overline{\kern -0.2em X}{}\xspace}

\def\BR         {{\ensuremath{\mathcal{B}}\xspace}}

\def\mes        {\mbox{$m_{\rm ES}$}\xspace}

\newcommand{\tev}{\ensuremath{\mathrm{\,Te\kern -0.1em V}}\xspace}
\newcommand{\gev}{\ensuremath{\mathrm{\,Ge\kern -0.1em V}}\xspace}
\newcommand{\mev}{\ensuremath{\mathrm{\,Me\kern -0.1em V}}\xspace}
\newcommand{\kev}{\ensuremath{\mathrm{\,ke\kern -0.1em V}}\xspace}
\newcommand{\ev}{\ensuremath{\mathrm{\,e\kern -0.1em V}}\xspace}
\newcommand{\gevc}{\ensuremath{{\mathrm{\,Ge\kern -0.1em V\!/}c}}\xspace}
\newcommand{\mevc}{\ensuremath{{\mathrm{\,Me\kern -0.1em V\!/}c}}\xspace}
\newcommand{\gevcc}{\ensuremath{{\mathrm{\,Ge\kern -0.1em V\!/}c^2}}\xspace}
\newcommand{\mevcc}{\ensuremath{{\mathrm{\,Me\kern -0.1em V\!/}c^2}}\xspace}

\def\invfb  {\ensuremath{\mbox{\,fb}^{-1}}\xspace}

\def\mus  {\ensuremath{\rm \,\mus}\xspace}

\def\mus        {\ensuremath{\,\mu{\rm s}}\xspace}    

\def\to                 {\ensuremath{\rightarrow}\xspace}

\def\pep2{PEP-II}

\def\gsim{{~\raise.15em\hbox{$>$}\kern-.85em
          \lower.35em\hbox{$\sim$}~}\xspace}
\def\lsim{{~\raise.15em\hbox{$<$}\kern-.85em
          \lower.35em\hbox{$\sim$}~}\xspace}

\xspace

\newcommand{\jprlBase}       {Phys.\ Rev.\ Lett.\xspace}
\newcommand{\jprBase}        {Phys.\ Rev.\xspace}
\newcommand{\jplBase}        {Phys.\ Lett.\xspace}
\newcommand{\nimBaseA}       {Nucl.\ Instr.\ Meth.\xspace}

\newcommand{\cpc}       [1]  {{Comput.\ Phys.\ Commun.\ {\bf #1}}}

\newcommand{\nima}      [1]  {\nimBaseA~A~{\bf #1}}

\newcommand{\plb}       [1]  {\jplBase\ B~{\bf #1}}

\newcommand{\jprd}      [1]  {\jprBase\ D~{\bf #1}}

\def\geant      {\mbox{\tt GEANT}\xspace}
\def\geant4     {\mbox{\tt GEANT4}\xspace}
\def\geantFour  {\mbox{\tt GEANT4}\xspace}

\def\jetset74   {\mbox{\tt Jetset \hspace{-0.5em}7.\hspace{-0.2em}4}\xspace}

%% file: pubboard/authors_may2007.tex
%
\author{B.~Aubert}
\author{M.~Bona}
\author{D.~Boutigny}
\author{Y.~Karyotakis}
\author{J.~P.~Lees}
\author{V.~Poireau}
\author{X.~Prudent}
\author{V.~Tisserand}
\author{A.~Zghiche}
\affiliation{Laboratoire de Physique des Particules, IN2P3/CNRS et Universit\'e de Savoie, F-74941 Annecy-Le-Vieux, France }
\author{J.~Garra~Tico}
\author{E.~Grauges}
\affiliation{Universitat de Barcelona, Facultat de Fisica, Departament ECM, E-08028 Barcelona, Spain }
% ...
\collaboration{The \babar\ Collaboration}
\noaffiliation

%% file: pubboard/acknow_PRL.tex
We are grateful for the excellent luminosity and machine conditions
provided by our \pep2\ colleagues, 
and for the substantial dedicated effort from
the computing organizations that support \babar.
The collaborating institutions wish to thank 
SLAC for its support and kind hospitality. 
This work is supported by
DOE
and NSF (USA),
NSERC (Canada),
CEA and
CNRS-IN2P3
(France),
BMBF and DFG
(Germany),
INFN (Italy),
FOM (The Netherlands),
NFR (Norway),
MIST (Russia),
MEC (Spain), and
STFC (United Kingdom). 
Individuals have received support from the
Marie Curie EIF (European Union) and
the A.~P.~Sloan Foundation.

%% file: paper.bbl
\begin{thebibliography}{99}
\bibitem{bib:cc}
Throughout the Letter the charge-conjugate modes are included implicitly.
\end{thebibliography}
